\documentclass[12pt,onecolumn]{IEEEtran}

\renewcommand{\baselinestretch}{1.8}

\usepackage{amsmath}
\usepackage{amsfonts}%
\usepackage{amssymb}%
\usepackage{graphicx}
\begin{document}

\title{A low-cost time-hopping impulse radio system for high data rate
transmission$^{1}$\thanks{$^{1}$Part of this work was presented at WPMC 2003,
Yokosuta, Japan.}}
\author{Andreas F. Molisch,$^{2}$\thanks{$^{2}$Author for correspondence. Email:
Andreas.Molisch@ieee.org.} Ye Geoffrey Li, Yves-Paul Nakache, Philip Orlik,
\and Makoto Miyake, Yunnan Wu, Sinan Gezici, Harry Sheng,
\and S. Y. Kung, H. Kobayashi, H.Vincent Poor, Alexander Haimovich$,$and Jinyun
Zhang$^{3}$\thanks{$^{3}$A. F. Molisch, Y. P. Nakache, P. Orlik, and J. Zhang
are with Mitsubishi Electric Research Labs, Cambridge, MA, USA. Y. G. Li is
with Georgia Institute of Technology, Atlanta, GA, USA. M. Miyake is with
Information Technology R\&D Center, Mitsubishi Electric Corp., Ofuna,
Japan..Y. Wu, S. Gezici, S. Y. Kung, H. Kobayashi, and H. V. Poor are with
Princeton University, Princeton, NJ, USA. H. Sheng and A. Haimovich are with
New Jersey Institute of Technology, Newark, NJ, USA.}}
\maketitle

\begin{abstract}
We present an efficient, low-cost implementation of time-hopping impulse radio
that fulfills the spectral mask mandated by the FCC and is suitable for
high-data-rate, short-range communications. Key features are: (i) all-baseband
implementation that obviates the need for passband components, (ii)
symbol-rate (not chip rate) sampling, A/D conversion, and digital signal
processing, (iii) fast acquisition due to novel search algorithms, (iv)
spectral shaping that can be adapted to accommodate different spectrum
regulations and interference environments. Computer simulations show that this
system can provide 110Mbit/s at 7-10m distance, as well as higher data rates
at shorter distances under FCC emissions limits. Due to the spreading concept
of time-hopping impulse radio, the system can sustain multiple simultaneous
users, and can suppress narrowband interference effectively.

\end{abstract}

\section{Introduction}

Ultrawideband (UWB) wireless systems are defined as systems that use either a
large relative bandwidth (ratio of bandwidth to carrier frequency larger than
25{\%}), or a large absolute bandwidth (larger than 500MHz). While UWB radar
systems have been used for a long time, mainly in the military domain
\cite{Taylor_1995}, UWB communications systems are a fairly recent
development. The first papers in the open literature are those of Win and
Scholtz \cite{Scholtz_1993}, \cite{Win_and_Scholtz_1998},
\cite{Win_and_Scholtz_2000}, who developed the concept of time-hopping impulse
radio (TH-IR) system. This concept excited immense interest in the area of
military \cite{Kolenchery_et_al_1998} as well as civilian \cite{Ho_et_al_2001}
communications. Further advances of TH-IR are described, e.g., in
\cite{LeMatret_and_Giannakis_2000}, \cite{Conroy_et_al_1999},
\cite{Forouzan_et_al_2000}, \cite{Cassioli_et_al_2003},
\cite{Fischler_and_Poor_2002}. In 2002, the Federal Communications Commission
(FCC) in the US allowed \textit{unlicensed} UWB communications \cite{FCC_2002}%
. This greatly increased commercial interest in UWB, leading to a large number
of papers, see, e.g., \cite{UWBST_2002}, \cite{Oulu_2003}.

One of the most promising applications is data communications at rates that
are higher than the currently popular 802.11b ($11$ Mbit/s) and 802.11a (%
%TCIMACRO{\TEXTsymbol{<}}%
%BeginExpansion
$<$%
%EndExpansion
$54$ Mbit/s) standards. The goal, as mandated, e.g., by the standardization
committee IEEE 802.15.3a, is a system that can provide multiple piconets with
$110$ Mbit/s each. This data rate should be achieved for distances up to $10$
m (Personal Area Networks). Higher data rates should be feasible at shorter distances.

The principle of using very large bandwidths has several generic advantages:

\begin{itemize}
\item By spreading the information over a large bandwidth, the spectral
\textit{density} of the transmit signal can be made very low. This decreases
the probability of intercept (for military communications), as well as the
interference to narrowband victim receivers.

\item The spreading over a large bandwidth increases the immunity to
narrowband interference and ensures good multiple-access capabilities
\cite{Zhao_and_Haimovich_2002}, \cite{Ramirez_2001}.

\item The fine time resolution implies high temporal diversity, which can be
used to mitigate the detrimental effects of fading
\cite{Win_and_Scholtz_1998_energy_capture}.

\item Propagation conditions can be different for the different frequency
components. For example, a wall might be more transparent in a certain
frequency range. The large bandwidth increases the chances that at least some
frequency components arrive at the receiver \cite{Cassioli_et_al_2002}.
\end{itemize}

These advantages are inherent in the use of very large bandwidths, and can
thus be achieved by \textit{any} UWB system, including the recently proposed
UWB frequency-hopping OFDM system \cite{Batra_et_al_2003} and UWB
direct-sequence spread spectrum (DS-SS) systems \cite{McCorkle_et_al_2003}.
However, TH-IR has additional advantages:

\begin{itemize}
\item Recent information-theoretic results indicate that higher capacities can
be achieved than with DS-SS systems \cite{Subramanian_and_Hajek_2002},
\cite{Molisch_et_al_2003_PACRIM}.

\item More important from a practical point of view, impulse radio systems
operate in baseband only, thus requiring no frequency upconversion circuitry
and associated RF components \cite{LeMatret_and_Giannakis_2000}, though
circuitry for accurate timing is still required. This allows low-cost implementation.
\end{itemize}

A lot of progress has been made in the theoretical understanding of impulse
radio, as evidenced by the papers mentioned above. However, several
assumptions made in the theoretical analyses do not agree with the
requirements for a practical implementation of a high-data-rate impulse radio
system. Those requirements may stem from the regulations by the FCC and other
frequency regulators, from the necessity of coexistence with other devices,
and from cost considerations. The goal of this paper is to describe the
complete physical-layer design of an IR system that is suitable for practical
implementation. In this system, we combine existing and innovative aspects,
giving special attention to the interplay between the different aspects. The
current paper is thus more of an \textquotedblleft
engineering\textquotedblright\ paper, while the theoretical background of some
of our innovations is described in \cite{Wu_et_al_2003},
\cite{Nakache_and_Molisch_2003}, \cite{Gezici_et_al_2003}.

The remainder of the paper is organized the following way: in Section II, we
present an overview of the system. Next, we discuss the transmit signal, and
how its spectrum can be shaped to fit the requirements of regulators, as well
as to minimize interference to nearby devices. Section IV describes the signal
detection at the receiver, including the structure of the Rake receiver and
the equalizer. The channel estimation procedure that is used for establishing
the weights of the Rake receiver and equalizer is discussed in Section V.
Finally, Section VI presents simulations of the total performance of the
system in terms of coverage and resistance to interference from narrowband
signals and other UWB transmitters. A summary and conclusions wrap up the paper.

\section{System overview}

\begin{figure}
[ptb]
\begin{center}
\includegraphics[
trim=0.000000in 0.000000in 0.000000in -0.596115in,
height=3.0116cm,
width=8.0309cm
]%
{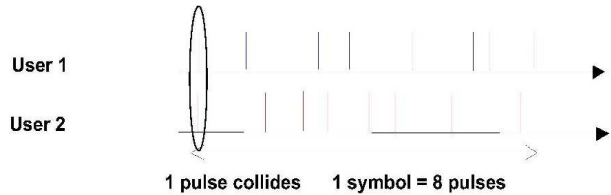}%
\caption{Principle of time-hopping impulse radio for the suppression of
catastrophic collisions.}%
\label{thir_principle}%
\end{center}
\end{figure}

The system that we are considering is a time-hopping impulse radio (TH-IR)
system. We first describe "classical" TH-IR \cite{Win_and_Scholtz_2000}. Each
data bit is represented by several short pulses; the duration of the pulses
determines essentially the bandwidth of the (spread) system. For the
single-user case, it would be sufficient to transmit a single pulse per
symbol. However, in order to achieve good multiple access (MA) properties, we
have to transmit a whole sequence of pulses. Since the UWB transceivers are
unsynchronized, so-called \textquotedblleft catastrophic
collisions\textquotedblright\ can occur, where pulses from several
transmitters arrive at the receiver almost simultaneously. If only a single
pulse would represent one symbol, this would lead to a bad
signal-to-interference ratio, and thus to high bit error probability BER.
These catastrophic collisions are avoided by sending a whole sequence of
pulses instead of a single pulse. The transmitted pulse sequence is different
for each user, according to a so-called time-hopping (TH) code. Thus, even if
one pulse within a symbol collides with a signal component from another user,
other pulses in the sequence will not. This achieves interference suppression
gain that is equal to the number of pulses in the system. Fig.
\ref{thir_principle} shows the operating principle of a generic TH-IR system.
We see that the possible positions of the pulses within a symbol follow
certain rules: the symbol duration is subdivided into $N_{f}$
\textquotedblleft frames\textquotedblright\ of equal length. Within each frame
the pulse can occupy an almost arbitrary position (determined by the
time-hopping code). Typically, the frame is subdivided into \textquotedblleft
chips\textquotedblright, whose length is equal to a pulse duration. The
(digital) time-hopping code now determines which of the possible positions the
pulse actually occupies.
%TCIMACRO{\FRAME{ftbpFU}{8.0309cm}{3.0116cm}{0pt}{\Qcb{Principle of
%time-hopping impulse radio for the suppression of catastrophic collisions.}%
%}{\Qlb{thir_principle}}{thir_principle.eps}%
%width 8.0309cm;  height 3.0116cm;  depth 0pt;  original-width 6.3053in;
%original-height 1.7339in;  cropleft "0";  croptop "1.3438";  cropright "1";
%filename 'figures/THIR_principle.eps';file-properties "XNPEU";}}}%
%BeginExpansion
%EndExpansion
The modulation of this sequence of pulses can be pulse-position modulation
(PPM), as suggested in \cite{Win_and_Scholtz_2000}, or amplitude modulation
(PAM). PPM has the advantage that the detector can be simpler (an energy
detector) in AWGN channels. However, reception in multipath environments
requires a Rake receiver for either PPM or PAM.

While this scheme shows good performance for some applications, it does have
problems for high-data rate, FCC-compliant systems:

\begin{enumerate}
\item Due to the use of PPM, the transmit spectrum shows spectral lines. This
requires the reduction of the total emission power, in order to allow the
fulfillment of the FCC mask within each 1MHz band, as required by the FCC.

\item Due to the high data rate required by 802.15, and due to the high delay
spread seen by indoor channels, the system works better with an equalizer. An
equalizer for PPM will introduce increased complexity and cost.

\item For a full recovery of all considered multi-path components, the system
requires a Rake receiver with a large number of fingers. A conventional
implementation, using many digital correlators, will also introduce increased
complexity and cost.

\item Due to the relatively low spreading factor of less than 40, the number
of possible pulse positions within a frame is limited. This might lead to
higher collision probability, and thus smaller interference suppression.
\end{enumerate}

The first two problems are solved by using (antipodal) pulse amplitude
modulation (PAM) instead of PPM. This eliminates the spectral lines, and
allows in general an easier shaping of the spectrum. Furthermore, it allows
the use of simple linear equalizers. As detailed below, an innovative Rake
receiver is considered to overcome the third problem; this Rake structure
implements correlators by means of pulse generators and multipliers only. The
problem of multiple-access interference, finally, can be addressed by
interference-suppressing combining of the Rake finger signals.
%TCIMACRO{\FRAME{ftbpFU}{15.0579cm}{10.0364cm}{0pt}{\Qcb{Blockdiagram of the
%transmitter (a) and receiver (b).}}{\Qlb{blockdiagram}}{blockdiagram.eps}%
%width 15.0579cm;  height 10.0364cm;  depth 0pt;  original-width 8.1699in;
%original-height 7.4556in;  cropleft "0";  croptop "1";  cropright "1.3719";
%cropbottom "0";  filename 'figures/blockdiagram.eps';file-properties "XNPEU";}%
%}}%
%BeginExpansion
\begin{figure}
[ptb]
\begin{center}
\includegraphics[
trim=0.000000in 0.000000in -3.038385in 0.000000in,
height=10.0364cm,
width=15.0579cm
]%
{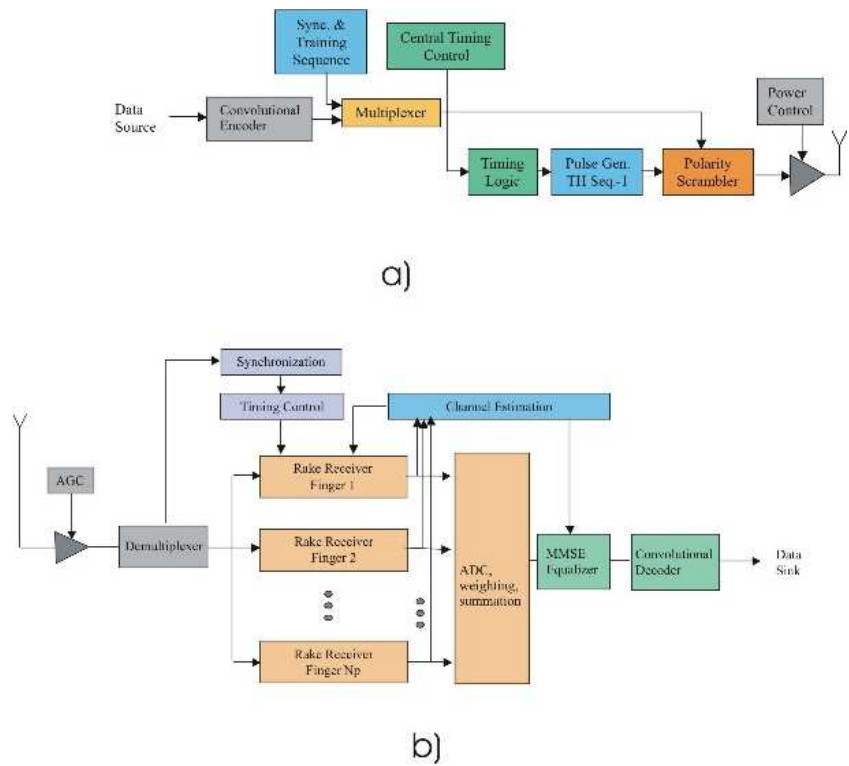}%
\caption{Blockdiagram of the transmitter (a) and receiver (b).}%
\label{blockdiagram}%
\end{center}
\end{figure}
%EndExpansion

A block diagram of the system is shown in Fig. \ref{blockdiagram}. The
transmit data stream is divided into blocks, and each block is encoded with a
convolutional coder. We use a rate $1/2$ convolutional code with a constraint
length 7. The use of turbo codes or low-density parity check codes would
improve the performance by approximately 2 dB; however, decoding becomes
challenging at the high data rates envisioned in this scheme. Then, a preamble
is prepended that can be used for both acquisition and channel estimation. As
mentioned above, the modulation and multiple access format is BPSK-modulated
TH-IR. Each pulse sequence representing one symbol is multiplied by $\pm$1,
depending on the bit to be transmitted. Finally, each data block (including
preamble) is amplified (with power control, in order to minimize interference
to other systems), and transmitted. Note that as the system is packet based
and the number of packets per second can vary, it is not desirable to code
across packets.

In the receiver, the acquisition part of the preamble is stripped off and used
to determine the timing. Once this has been established, the \textquotedblleft
channel estimation part\textquotedblright\ of the preamble is used to
determine the coefficients for the Rake receiver and the equalizer. The main
body of the data block is then received by a Rake receiver that can be
interpreted as a filter that is matched to the convolution of the transmit
signal and the channel impulse response. Each finger of the Rake finger is a
filter that is matched to a time-delayed version of the transmit signal,
encompassing both to the pulse shape and the time-hopping sequence. We use
here an innovative Rake structure that requires only pulse generators and no
delays, which makes an analogue implementation possible -- this allows us to
perform the sampling and A/D conversion only at the \emph{symbol rate},
instead of the chip rate. Note that for chip rate sampling, A/D converters
with about $20$ Gsamples/s would be required. The outputs of the Rake fingers
are weighted (according to the principles of optimum combining) and summed up.
The optimum location and weight of the fingers can be determined from the
channel sounding sequence, which is processed before the reception of the
actual data. The output of the summer is then sent through an MMSE equalizer
and a decoder for the convolutional code.

One important point of the system is that all the pulses are \textit{baseband}
pulses, more specifically, derivatives of Gaussian pulses. This allows a
simple pulse generation, and obviates any need for passband components. This
is a typical property of time-hopping impulse radio; however, it is not a
trivial task within the restrictions of the FCC that the main power is emitted
in the $3-10$ GHz range. We will show in Sec. 3 how this can be achieved.

The goal of our design is to obtain a low-cost implementation. Thus, the
design is not theoretically optimum, but rather contains a number of
simplifications that reduce complexity of implementation and costs.

\section{Transmit signal and spectral shaping}

\subsection{Mathematical description of the transmit signal}

Throughout this paper, we use a communication system model where the
transmitted signal is given by
\begin{equation}
s_{tr}(t)=\sum\limits_{j=-\infty}^{\infty}d_{j}b_{\left\lfloor j/N_{f}%
\right\rfloor }w_{\text{tr}}(t-jT_{f}-c_{j}T_{c})=\sum\limits_{k=-\infty
}^{\infty}b_{k}w_{\text{seq}}\left(  t-kT_{s}\right)
\end{equation}
where $w_{\text{tr}}(t)$ is the transmitted unit-energy pulse, $T_{f}$ is the
average pulse repetition time, $N_{f}$ is the number of frames (and therefore
also the number of pulses) representing one information symbol of length
$T_{s}$, and $b$ is the information symbol transmitted, i.e., $\pm1;$
$w_{\text{seq}}(t)$ is the pulse sequence transmitted representing one symbol.
The TH sequence provides an additional time shift of $c_{j}T_{c}$ seconds to
the $j^{th}$ pulse of the signal, where $T_{c}$ is the chip interval, and
$c_{j}$ are the elements of a pseudorandom sequence, taking on integer values
between $0$ and $N_{c}-1$. To prevent pulses from overlapping, the chip
interval is selected to satisfy $T_{c}\leq T_{f}$/$N_{c}$; in the following,
we assume $T_{f}$ / $T_{c}=N_{c}$ so that $N_{c}$ is the number of chips per
frame. We also allow \textquotedblleft polarity scrambling\textquotedblright%
\ (see Sec. III.4), where each pulse is multiplied by a (pseudo) random
variable $d_{j}$ that can take on the values $+1$ or $-1$ with equal
probability. The sequence $d_{j}$ is assumed to be known at transmitter and receiver.

An alternative representation can be obtained by defining a sequence
{\{}$s_{j}${\}} as follows
\begin{equation}
s_{j}=\left\{  {%
\begin{array}
[c]{l}%
d_{\left\lfloor {j/N_{c}}\right\rfloor }\text{ \ \ \ \ \ \ for \ \ \ \ \ \ \ }%
j-N_{f}\left\lfloor {j/N_{c}}\right\rfloor \text{=c}_{\left\lfloor {j/N_{c}%
}\right\rfloor }\\
0\text{ \ \ \ \ \ \ \ \ \ \ \ \ \ \ otherwise}%
\ \ \ \ \ \ \ \ \text{\ \ \ \ \ }%
\end{array}
}\right. \label{eq1}%
\end{equation}
Then the transmit signal can be expressed as
\begin{equation}
s_{tr}(t)=\sum\limits_{j=-\infty}^{\infty}s_{j}b_{\left\lfloor {j/N_{f}N_{c}%
}\right\rfloor }w_{\text{tr}}(t-jT_{c}).\label{eq2}%
\end{equation}

To satisfy the spectrum masking requirement of the FCC, the transmit waveform
$w_{tr}$, also known as monocycle waveform, is chosen to be the 5$^{th}$
derivative of the Gaussian pulse and it can be expressed as,
\begin{equation}
w_{\text{tr}}(t)=p\left(  t\right)  =K_{2}\left(  {-15\frac{t}{\sigma_{p}%
}+10\frac{t^{3}}{\sigma_{p}^{3}}-\frac{t^{5}}{\sigma_{p}^{5}}}\right)
\exp(-\frac{t^{2}}{2\sigma_{p}^{2}}),
\end{equation}
where $K_{2}$ is a normalization constant, and $\sigma_{p}$ controls the width
of the pulse and it is chosen according to the spectral mask requirement of
the FCC, which is \cite{Sheng_et_al_2003},%
\begin{equation}
\sigma_{p}=5.08\times10^{-11}\text{ s}.
\end{equation}
Other signals shapes are possible; in particular, a combination of weighted
pulses $p(t)$ (as explained below) can be used to improve the spectral
properties. The various methods (e.g., Rake receiver, pulse polarity
randomization, ....) discussed in the remainder of the paper can be applied
independently of the exact shape of the transmit waveform.

\subsection{Spectral shaping - general aspects}

One of the key requirements for a UWB system is the fulfillment of the
emission mask mandated by the national spectrum regulators
\cite{Lehman_and_Haimovich_2003a}. In the USA, this mask has been prescribed
by the FCC and essentially allows emissions in the $3.1-10.6$ GHz range with
power spectral density of $-41.3$ dBm/MHz; in Europe and Japan, it is still
under discussion. In addition, emissions in certain parts of the band
(especially the $5.2-5.8$ GHz range used by wireless LANs) should be kept low,
as UWB transceivers and IEEE 802.11a transceivers, which operate in the $5$
GHz range, are expected to work in close proximity. We are using two
techniques in order to fulfill those requirements.

\begin{itemize}
\item The first is a linear combination of a set of basis pulses to be used
for shaping of the spectrum of a transmitted impulse radio signal. The delayed
pulses are obtained from several appropriately timed programmable pulse
generators. The computation of the delays and weights of those pulses is
obtained in a two-step optimization procedure \cite{Wu_et_al_2003}.

\item A further improvement of the spectral properties can be obtained by
exploiting different polarities of the pulses that constitute a transmit
sequence $w_{\operatorname{seq}}(t)$. Using different pulse polarities does
not change anything for the signal detection, as it is known at the receiver,
and can thus be easily reversed. However, it does change the spectrum of the
\emph{emitted} signal, and thus allows a better matching to the desired
frequency mask \cite{Nakache_and_Molisch_2003},
\cite{Lehmann_and_Haimovich_2003}.
\end{itemize}

The first technique (combination of pulses) leads to a shaping of the
spectrum, allowing the placement of broad minima and an efficient
\textquotedblleft filling out\textquotedblright\ of the FCC mask. The second
technique is used to reduce or eliminate the peak-to-average ratio of the
spectrum, and allows the design of more efficient multiple-access codes. Note
that these two aspects are interrelated, and the optimization of pulse
combination and polarity randomization should be done jointly in order to
achieve optimality. However, such a joint treatment is usually too complicated
for adaptive modifications of the transmit spectrum.

A further important aspect of the spectral shaping is that it can be used not
only to reduce interference $to$ other devices, but also interference $from$
narrowband interferers. This can be immediately seen from the fact that
matched filtering is used in the receiver. Placing a null in the transmit
spectrum thus also means that the receiver suppresses this frequency.
Furthermore, it might be advantageous in some cases to perform "mismatched
filtering" at the receiver by placing minima in the receive transfer function
even if there is no corresponding minimum in the transmit spectrum. This is
useful especially for the suppression of narrowband interferers that could
otherwise drive the A/D converter into saturation.

\subsection{Pulse combination}

\begin{figure}
[ptb]
\begin{center}
\includegraphics[
trim=0.000000in 0.000000in -2.668759in 0.000000in,
height=6.0231cm,
width=8.0309cm
]%
{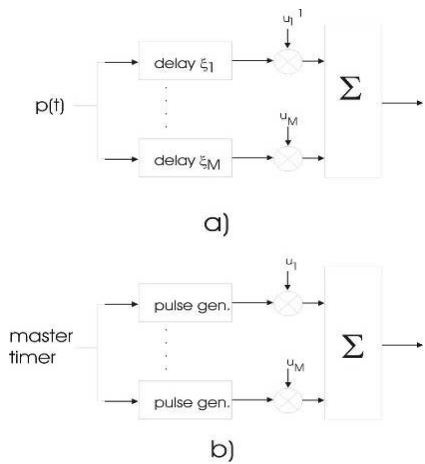}%
\caption{Principle of pulse combination for spectral shaping with delay lines
(a) and with programmable pulse generators (b).}%
\label{pulsecombination}%
\end{center}
\end{figure}

One of the key problems of \textquotedblleft conventional\textquotedblright%
\ TH-IR radio is that it is difficult to influence its spectrum without the
use of RF components. Spectral notches, e.g., are typically realized by means
of bandblock filters. However, this is undesirable for extremely low cost
applications; furthermore, it does not allow adaptation to specific
interference situations. We have thus devised a new scheme for shaping the
spectrum \cite{Wu_et_al_2003}. This scheme uses delaying and weighting of a
set of basis pulses to influence the transmit spectrum, see Fig.
\ref{pulsecombination}.%
%TCIMACRO{\FRAME{ftbpFU}{8.0309cm}{6.0231cm}{0pt}{\Qcb{Principle of pulse
%combination for spectral shaping with delay lines (a) and with programmable
%pulse generators (b).}}{\Qlb{pulsecombination}}{pulsecombination.eps}%
%width 8.0309cm;  height 6.0231cm;  depth 0pt;  original-width 4.0534in;
%original-height 2.4915in;  cropleft "0";  croptop "1";  cropright "1.6584";
%filename 'figures/pulsecombination.eps';file-properties "XNPEU";}}}%
%BeginExpansion
%EndExpansion

The basic transmit waveform $w_{\text{tr}}(t)$ is a sum of delayed and
weighted "basic pulse shapes" $p(t)$ that can be easily generated, e.g.,
Gaussian pulses and their derivatives.
\begin{align}
w_{tr}(t) &  \equiv\sum_{i=0}^{M}u{_{i}}p(t-\xi_{i})\\
W(\text{j}\Omega) &  \equiv\int_{-\infty}^{\infty}w_{tr}(t)e^{-j\Omega
t}dt=\sum_{i=0}^{M}u{_{i}}P(j\Omega)e^{-j\Omega\xi_{i}},\label{eq:S_Omega}%
\end{align}
where j is the imaginary unit (not to be confused with the index $j$ that
denotes the considered frame), $u_{i}$ are the pulse weights, $W($j$\Omega)$
is the Fourier transform of $w_{\text{tr}}(t)$, and $\Omega$ is the transform
variable. In contrast to tapped delay lines, where only certain discrete
delays are feasible, we assume here that a continuum of delays can be chosen.
This can be achieved by the use of programmable pulse generators. The range of
allowed delays of the coefficients is determined by the pulse repetition
frequency of the communication system. The number of pulse generators $M+1$
should be kept as low as possible to reduce the implementation costs.

Let us introduce the following notations:
\begin{align}
&  \underline{u}\equiv\lbrack u_{0}~u_{1}~\ldots~u_{M}]^{T}\label{eq:p}\\
&  \underline{\xi}\equiv\lbrack\xi_{0}~\xi_{1}~\ldots~\xi_{M}]^{T}\\
&  r(\lambda)\equiv\int_{-\infty}^{\infty}p(t-\lambda)p(t)dt=r(-\lambda),\\
&  \mathbf{R}(\underline{\xi})\equiv\left(
\begin{array}
[c]{cccc}%
r(0) & r(\xi_{0}-\xi_{1}) & \cdots & r(\xi_{0}-\xi_{M})\\
r(\xi_{1}-\xi_{0}) & r(0) & \ddots & r(\xi_{1}-\xi_{M})\\
\vdots & \ddots & \ddots & \vdots\\
r(\xi_{M}-\xi_{0}) & r(\xi_{M}-\xi_{1}) & \cdots & r(0)
\end{array}
\right)  ,\\
&  \langle w_{tr}(t),w_{tr}(t)\rangle\equiv\int_{-\infty}^{\infty}%
w_{tr}(t)w_{tr}(t)dt=\underline{u}^{T}\mathbf{R}(\underline{\xi})\underline{u}%
\end{align}

The single user spectrum shaping problem can now be formulated as follows:
\begin{equation}
\max_{\underline{u},\underline{\xi}}\langle w_{tr}(t),w_{tr}(t)\rangle
,\;\text{subject to}\;|W(j\Omega)|^{2}\leq M(\Omega)\;,\forall\Omega\in
\lbrack-\infty,\infty],
\end{equation}
where $M(\Omega)$ is the upper-bound on the magnitude response regulated by
FCC. This is equivalent to:
\begin{equation}
\min_{\underline{u},\underline{\xi}}\max_{\Omega\in\lbrack-\infty,\infty
]}\frac{|W(j\Omega)|^{2}}{M(\Omega)},\qquad\text{subject to}\;\;\underline
{u}^{H}R(\underline{\xi})\underline{u}=1.
\end{equation}

The criteria for the optimization $M(\Omega)$ can thus stem from the FCC
spectral mask, which is fixed, from the necessity to avoid interference to
other users, which can be pre-defined or time-varying, or following an
instantaneous or averaged determination of the emissions of users in the
current environment, or other criteria. In any case, these criteria are mapped
onto an \textquotedblleft instantaneous" spectral mask that has to be
satisfied by the pulse. If the fulfillment of the FCC spectral mask is the
only requirement, then the optimum weights can be computed a priori, and
stored in the transceivers; in that case, the computation time determining the
optimum weights and delays is not relevant, and exhaustive search can be used.
However, in order to adjust to different interference environments, a
capability to optimize the weights dynamically is desirable. This can be
achieved, e.g., by an efficient two-step procedure that in the first step uses
an \emph{approximate} formulation of the optimization problem, namely 2-norm
minimization that can be solved in closed form. This solution is then used as
the initialization of a nonlinear optimization (e.g., by means of a neural
network) to find the solution to the \emph{exact} formulation. Details of this
two -step procedure can be found in \cite{Wu_et_al_2003}. Note also that the
spectral shaping can be refined even more by combining different basis pulses.
However, this requires different pulse generators, which increases
implementation complexity.

\subsection{Polarity randomization}

\begin{figure}
[ptb]
\begin{center}
\includegraphics[
trim=0.000000in 0.000000in -2.638572in 0.000000in,
height=8.0309cm,
width=15.0579cm
]%
{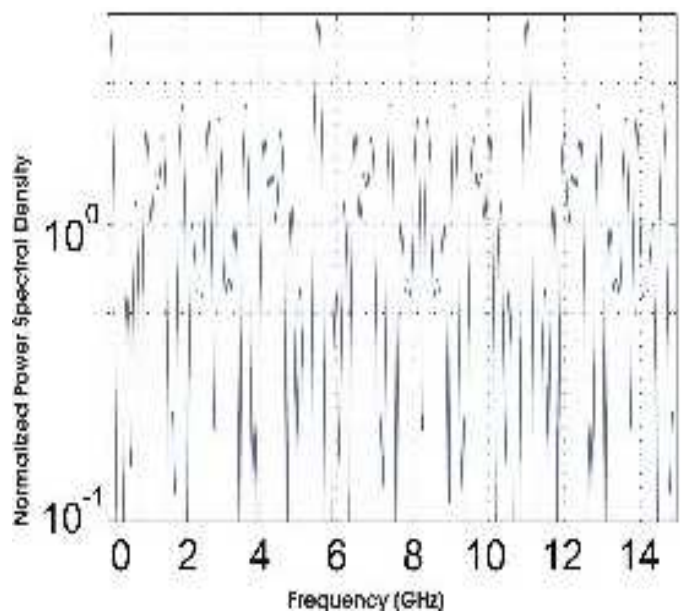}%
\caption{Spectrum of time hopping sequence with "classical" TH sequence. 5
chips each in 5 frames. Positions of the pulses given by the chip sequence [1
0 0 0 0 0 0 0 1 0 0 0 1 0 0 0 0 1 0 0 0 0 0 1 0].}%
\label{spectrum_ripples}%
\end{center}
\end{figure}

\begin{figure}
[ptbptb]
\begin{center}
\includegraphics[
trim=0.000000in 0.000000in -2.638572in 0.000000in,
height=8.0309cm,
width=15.0579cm
]%
{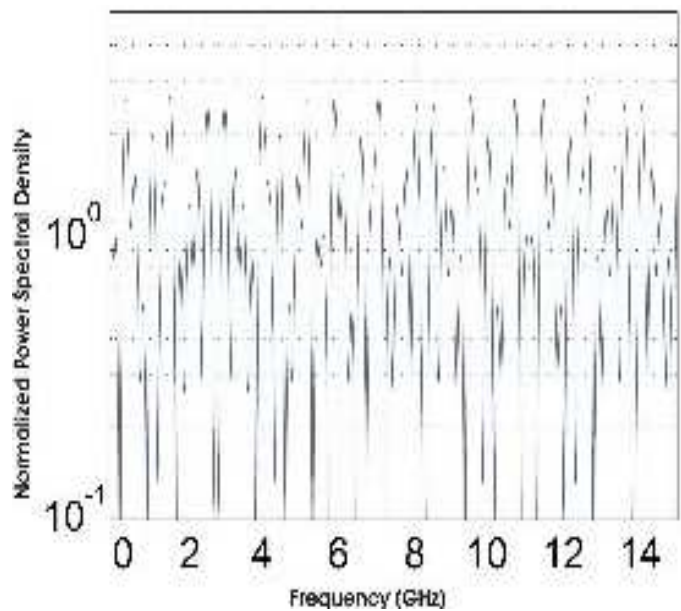}%
\caption{Spectrum of PAM signal with polarity randomization of the TH
sequence. Same positions of the pulses as in Fig. 4, with normalized pulse
amplitudes given by the weighting vector [-0.5 0.5 0.5 -0.5 1.5].}%
\label{spectrum_smoothed}%
\end{center}
\end{figure}

Conventional impulse radio systems use only a pseudo-random variation of the
pulse position to distinguish between different users. For PAM - TH-IR, the
spectrum of the transmit signal is determined by the spectrum of the transmit
waveform $w_{tr}(t)$, multiplied with the spectrum of the TH sequence. Fig.
\ref{spectrum_ripples} shows an example of a spectrum with a short (4 frames)
time hopping sequence, in combination with a 5th-order Gaussian basis pulse.
We can observe strong ripples, so that the peak-to-average ratio is about 6dB.
However, the ideal case would be to find TH sequences whose spectrum is flat,
so that the we can design the transmit waveform to fit the spectral mask as
closely as possible. One way to achieve this goal is to use very long TH
sequences (much longer than a symbol duration). However, this complicates the
design of the receiver, especially the equalizer. Alternatively, we can use
more degrees of freedom in the design of short sequences by allowing different
amplitudes and polarities of the pulses for the design of the sequence. This
helps to limit the power back-off by reducing the peak to average ratio.
However, it is still true that the less pulses compose the sequence, the
larger is the peak-to-average ratio. An example can be seen in Figs.
\ref{spectrum_ripples} (unipolar sequence) and Fig. \ref{spectrum_smoothed}
(polarity randomization); it is obvious that the ripples have been
considerably reduced; specifically, we reduced the peak-to-average ratio by
1.6 dB. We also have to bear in mind that we need to generate a multitude of
sequences that all should have the desired spectral properties, as well as
approximate orthogonality with respect to each other for arbitrary time shifts
of the sequences. This is a complex optimization problem, and has to be solved
by an exhaustive search.
%TCIMACRO{\FRAME{ftbpFU}{15.0579cm}{8.0309cm}{0pt}{\Qcb{Spectrum of time
%hopping sequence with "classical" TH sequence. 5 chips each in 5 frames.
%Positions of the pulses given by the chip sequence [1 0 0 0 0 0 0 0 1 0 0 0 1
%0 0 0 0 1 0 0 0 0 0 1 0].}}{\Qlb{spectrum_ripples}}{ripplespectrum.eps}%
%width 15.0579cm;  height 8.0309cm;  depth 0pt;  original-width 5.2909in;
%original-height 3.173in;  cropleft "0";  croptop "1";  cropright "1.4987";
%filename 'figures/ripplespectrum.eps';file-properties "XNPEU";}}}%
%BeginExpansion
%EndExpansion%
%TCIMACRO{\FRAME{ftbpFU}{15.0579cm}{8.0309cm}{0pt}{\Qcb{Spectrum of PAM signal
%with polarity randomization of the TH sequence. Same positions of the pulses
%as in Fig. 4, with normalized pulse amplitudes given by the weighting vector
%[-0.5 0.5 0.5 -0.5 1.5].}}{\Qlb{spectrum_smoothed}}{smoothedspectrum.eps}%
%width 15.0579cm;  height 8.0309cm;  depth 0pt;  original-width 5.2909in;
%original-height 3.173in;  cropleft "0";  croptop "1";  cropright "1.4987";
%filename 'figures/smoothedspectrum.eps';file-properties "XNPEU";}}}%
%BeginExpansion
%EndExpansion

\section{Signal detection\textbf{\ }}

\subsection{Received signal and Rake reception}

The Rake receiver is a key aspect of ultrawideband systems.\footnote{An
exception is OFDM-based UWB systems, which use a different principle to
collect the multipath energy \cite{Batra_et_al_2003}.} Due to the ultra wide
bandwidth, UWB systems have very fine temporal resolution, and are thus
capable of resolving multi-path components that are spaced approximately at an
inverse of the bandwidth. This is usually seen as a big advantage of UWB.
Multipath resolution of components reduces signal fading because the
multi-path components (MPCs) undergo different fading, and thus represent
different diversity paths. The probability that the components are
simultaneously all in a deep fade is very low. However, the fine time
resolution also means that many of the MPCs have to be \textquotedblleft
collected\textquotedblright\ by the Rake receiver in order to obtain all of
the available energy. A channel with $N_{\text{p}}$ resolvable paths requires
$N_{\text{p}}$ fingers to collect all of the available energy. In a dense
multi-path environment, the number of MPCs increases linearly with the
bandwidth. Even a sparse environment, such as specified by the IEEE 802.15.3a
standard channel model \cite{Molisch_et_al_2003_Mag}, requires up to 80
fingers to collect 80\% of the available energy.

Another problem is the complexity of the Rake fingers. In the conventional
Rake finger of a direct-sequence-spread spectrum (DS-SS) system, the received
signal is filtered with a filter matched to the chip waveform, and then in
each Rake finger, correlated to time-shifted versions of the spreading
sequence. In order to do the correlation, the signal first has to be sampled
and analog-to-digital (A/D) converted at the chip rate. Then, those samples
have to be processed. This involves convolution with the stored reference
waveform, addition, and readout. Sampling and A/D converting at the chip rate,
e.g., $10$ Gsamples/s, requires expensive components.\footnote{Note that some
companies have proposed the use of \emph{one-bit} A/D converters with $7.5-20$
Gsamples per second \cite{McCorkle_et_al_2003}.}

\begin{figure}
[ptb]
\begin{center}
\includegraphics[
trim=0.000000in 0.000000in -1.359689in 0.000000in,
height=6.0231cm,
width=15.0579cm
]%
{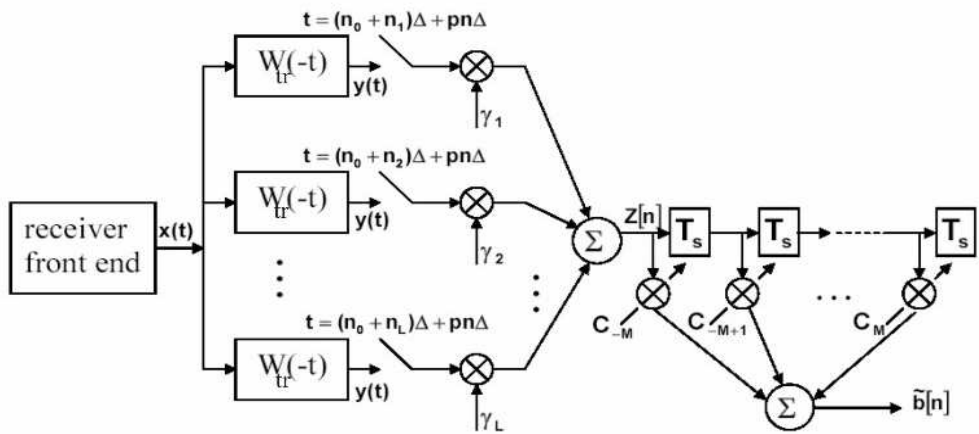}%
\caption{Structure of Rake receiver and equalizer.}%
\label{Rakestructure}%
\end{center}
\end{figure}

We avoid those problems by utilizing a Rake/equalizer structure as outlined in
Fig. \ref{Rakestructure}. Each Rake finger includes a programmable pulse
generator, controlled by a pulse sequence controller. The signal from the
pulse generator is multiplied with the received signal. The output of the
multiplier is then sent through a low-pass filter, which generates an output
proportional to a time integral of an input to the filter. The implementation
is analogue, while the adjustable delay blocks have been eliminated. The
hardware requirements for each Rake finger are: one pulse generator (which can
be controlled by the same timing controller), one multiplier, and one sampler
/ AD converter. It is an important feature of this structure that the sampling
occurs at the $symbol$ rate, not the chip rate. In the following, we assume
the use of $10$ Rake fingers; this is a very conservative number. Obviously, a
larger number of Rake fingers would give better performance; this is one of
the complexity/performance trade-offs in our design
\cite{Cassioli_et_al_2002_ICC}, \cite{Choi_and_Stark_2002}. The weights for
the combination of the fingers are determined by the channel estimation
procedure described in Sec. V.%
%TCIMACRO{\FRAME{ftbpFU}{15.0579cm}{6.0231cm}{0pt}{\Qcb{Structure of Rake
%receiver and equalizer.}}{\Qlb{Rakestructure}}{rakestructure.eps}%
%width 15.0579cm;  height 6.0231cm;  depth 0pt;  original-width 6.6521in;
%original-height 2.9447in;  cropleft "0";  croptop "1";  cropright "1.2044";
%filename 'figures/Rakestructure.eps';file-properties "XNPEU";}}}%
%BeginExpansion
%EndExpansion

Next, we compute the output of the different Rake fingers. Let the impulse
response of a UWB channel be
\begin{equation}
h\left(  t\right)  =\sum\limits_{k}\alpha_{k}\delta\left(  t-\tau_{k}\right)
,
\end{equation}
where $\tau_{k}$ and $\alpha_{k}$ are the delay and (real) gain of the k-th
path of the UWB channel, respectively. Then the channel output can be
expressed as
\begin{equation}
x\left(  t\right)  =h\left(  t\right)  \ast s_{tr}(t)+\overline{n}\left(
t\right)  =\sum\limits_{n=-\infty}^{\infty}b_{n}\widehat{h}\left(
t-nT_{s}\right)  +\overline{n}(t),\label{eq6}%
\end{equation}
where
\begin{equation}
\widehat{h}\left(  t\right)  =\sum\limits_{k}\alpha_{k}w_{\text{tr}}\left(
t-\tau_{k}\right)  .
\end{equation}
The output of the matched filter can be expressed as
\begin{equation}
y\left(  t\right)  =x\left(  t\right)  \ast w_{tr}\left(  {-t}\right)
=\sum\limits_{k=-\infty}^{\infty}b_{k}\tilde{h}\left(  t-kT_{s}\right)
+\tilde{n}\left(  t\right)  ,\label{eq7}%
\end{equation}
where
\begin{equation}
\tilde{h}\left(  t\right)  =\int\widehat{h}\left(  t-\tau\right)
w_{\text{tr}}\left(  {-\tau}\right)  d\tau=\sum\limits_{k}\alpha_{k}r\left(
t-\tau_{k}\right)  ,\label{eq8}%
\end{equation}%
\begin{equation}
r\left(  t\right)  =\int w_{\text{tr}}\left(  t+\tau\right)  w_{\text{tr}%
}\left(  \tau\right)  d{\tau,}%
\end{equation}
and
\begin{equation}
\mbox{ }\tilde{n}\left(  t\right)  =\overline{n}\left(  t\right)  \ast
w_{\text{tr}}\left(  {-t}\right)  .\label{eq9}%
\end{equation}

The samples of the matched filter output can be thus written as
\begin{equation}
y[n]=y\left(  {n\Delta}\right)  =\sum\limits_{k=-\infty}^{\infty}{b_{k}%
\tilde{h}\left(  {n\Delta-kp\Delta}\right)  }+\tilde{n}\left(  {n\Delta
}\right)  ,\label{eq10}%
\end{equation}
where $\Delta$ is the minimum time difference between Rake fingers and
$p=T_{s}/\Delta.$

\subsection{Combining of the Rake finger signals}

Let $\tilde{h}\left(  n_{l}\Delta\right)  $'s, for $l=1,{\ldots},L$ be the L
taps with the largest absolute values, $\left\vert \tilde{h}\left(
{n_{l}\Delta}\right)  \right\vert $'s. The output of the Rake receiver can be
expressed as
\begin{equation}
z\left[  n,n_{o}\right]  =\sum\limits_{l=1}^{L}\gamma_{l}y\left[
pn+n_{l}+n_{o}\right]  ,\label{eq11}%
\end{equation}
where $\gamma_{l}$ is the weight for the $l$-th finger and $n_{o}$ is a time
offset. It is obvious that the signal quality of the Rake receiver output
depends on the weight and initial time offset.

\emph{Maximal ratio combining (MRC)} is a traditional approach to determine
the weights of the Rake combiner. For the MRC Rake combiner, $\gamma
_{l}=\tilde{h}\left(  {n_{l}\Delta}\right)  $, and
\begin{equation}
z\left[  n,n_{o}\right]  =\sum\limits_{l=1}^{L}\tilde{h}\left(  n_{l}%
\Delta\right)  y\left[  pn+n_{l}+n_{o}\right]  .
\end{equation}
\textit{Minimum mean-square-error} (MMSE) Rake combining can improve the
performance of the Rake receiver in the presence of interference, including
intersymbol interference and multi-user interference since it automatically
take the correlation of the interference into consideration. For the MMSE Rake
combiner, the weights are determined to minimize
\begin{equation}
E\left\vert z[n,n_{o}]-b_{n}\right\vert ^{2}.
\end{equation}
The performance of the Rake receiver can be further improved if
\textit{adaptive timing} is used with the MMSE Rake combiner. That is, the
goal is to find optimum time offset $n_{o}$ and $\gamma_{l}$ to minimize
\begin{equation}
E\left\vert z[n,n_{o}]-b_{n}\right\vert ^{2}.
\end{equation}

When there is co-channel interference, the received signal can be written as%
\begin{equation}
\bar{y}[n]=\sum\limits_{k=-\infty}^{\infty}{b_{k}}\tilde{h}\left(
n\Delta-kT_{s}\right)  +\underbrace{\sum\limits_{k=-\infty}^{\infty}\bar
{b}_{k}\bar{h}\left(  n\Delta-kT_{s}\right)  +\tilde{n}\left(  n\Delta\right)
}_{i\left[  n\right]  },
\end{equation}
where $\{\bar{b}_{k}\}$ and $\bar{h}\left(  n\Delta-kT_{s}\right)  $ are,
respectively, i.i.d. sequence and channel impulse response corresponding to
the interferer, and $i[n]$ represents the interference-plus-noise. It can be
shown that $i\left[  n\right]  $ is not stationary but rather
cyclo-stationary. Let
\begin{equation}
P_{k}=E\left\{  \left\vert i\left[  mp+k\right]  \right\vert ^{2}\right\}  ,
\end{equation}
for any integer m and $k=0,1,{\ldots},p-1$. Therefore, for different $k$,
$h(nT_{s}+k\Delta)$ experiences different interference power. To improve the
performance of the Rake receiver, we need to normalize the channel impulse
response corresponding to the desired signal by
\begin{equation}
\hat{h}\left(  {n\Delta}\right)  =\frac{\tilde{h}\left(  {n\Delta}\right)
}{\sqrt{P_{k}}},
\end{equation}
and then find the $L$ taps with the largest absolute values of channel taps,
$\left\vert {\hat{h}\left(  {n_{l}\Delta}\right)  }\right\vert $'s for the
Rake receiver.

\begin{figure}
[ptb]
\begin{center}
\includegraphics[
trim=0.000000in 0.000000in -3.590082in 0.000000in,
height=6.0231cm,
width=15.0579cm
]%
{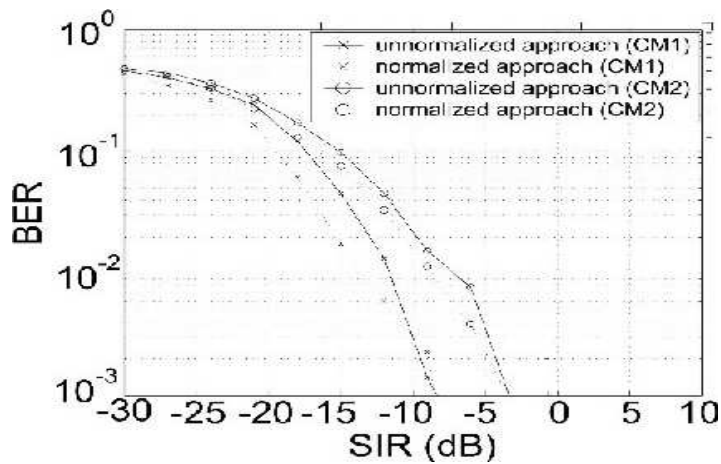}%
\caption{Interference suppression performance. One interferer, $SNR=50$ dB.}%
\label{interference_suppression}%
\end{center}
\end{figure}

Fig. \ref{interference_suppression} demonstrates the interference suppression
performance for a UWB system with one interferer and 50 dB SNR. We compare the
BER without normalization to the improved one that is normalized by noise
power as described above. Note that this can also be interpreted as the
difference between assuming the noise being stationary or cyclo-stationary.
%TCIMACRO{\FRAME{ftbpFU}{15.0579cm}{6.0231cm}{0pt}{\Qcb{Interference
%suppression performance. One interferer, $SNR=50$ dB.}}%
%{\Qlb{interference_suppression}}{interference_suppression.eps}%
%width 15.0579cm;  height 6.0231cm;  depth 0pt;  original-width 3.5708in;
%original-height 2.8323in;  cropleft "0";  croptop "1";  cropright "2.0054";
%filename 'figures/interference_suppression.eps';file-properties "XNPEU";}}}%
%BeginExpansion
%EndExpansion

\subsection{Channel equalizer}

The combination of the channel and the Rake receiver constitutes an equivalent
channel; however, since the symbol duration is shorter than the delay spread
of the channel, intersymbol interference (ISI) does occur. We combat that by
means of a MMSE (minimum mean square error) equalizer, as indicated in Fig.
\ref{Rakestructure}. The reasons for choosing a linear equalizer, instead of a
DFE, are twofold:

\begin{itemize}
\item the system is intended to operate at symbol error probabilities of
1-10\%; strong coding is used to decrease the frame error probability. Thus, a
decision feedback of the "raw symbols" (hard decision before the decoder)
would result in strong error propagation.

\item the alternative to use the symbols after decision would require
re-encoding and re-modulation before subtraction. This increases complexity
considerably. As the ISI is not a dominant source of errors in our system (as
determined from simulations that are not described in detail in Sec. VI), the
possible gains from this improved DFE scheme do not warrant such an increase
in complexity.
\end{itemize}

After the Rake receiver, a linear equalizer is used to mitigate residual
interference. Let the coefficients of the equalizer be $\left\{
c_{-K},c_{-K+1},....c_{-1},c_{0},c_{1},.....c_{K}\right\}  $. Then the
equalizer output is
\begin{equation}
\tilde{b}[n]=\sum\limits_{k=-K}^{K}c_{k}z\left[  n-k,n_{o}\right]  .
\end{equation}
To optimize performance, the equalizer coefficients are chosen to minimize the
MSE of its output, that is
\begin{equation}
MSE=E\left\vert \tilde{b}[n]-b_{n}\right\vert ^{2}.
\end{equation}
For the numerical simulations in Sec. VI, we will use a 5-tap equalizer.

\section{Parameter estimation}

\begin{figure}
[ptb]
\begin{center}
\includegraphics[
trim=0.000000in 0.000000in -0.481630in 0.000000in,
height=4.0154cm,
width=8.0309cm
]%
{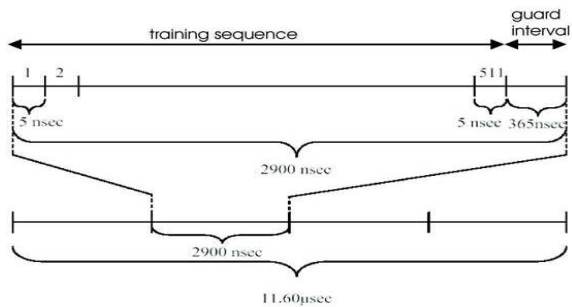}%
\caption{Structure of the training sequence.}%
\label{training_structure}%
\end{center}
\end{figure}

A training sequence is used to determine the parameters for the Rake receivers
and equalizers. It is desirable to use the correlators and A/D converters of
the Rake receivers, since these components have to be available anyway. This
is not straightforward, as the sampling and A/D conversion of the correlator
outputs is done at the symbol rate, while the channel parameters have to be
available for each possible chip sampling instant. This problem is solved by
combining a "sliding correlator" approach with a training sequence that
exhibits a special structure, as shown in Fig. \ref{training_structure}.%

%TCIMACRO{\FRAME{ftbpFU}{8.0309cm}{4.0154cm}{0pt}{\Qcb{Structure of the
%training sequence.}}{\Qlb{training_structure}}{training_structure.eps}%
%width 8.0309cm;  height 4.0154cm;  depth 0pt;  original-width 3.8164in;
%original-height 2.13in;  cropleft "0";  croptop "1";  cropright "1.1262";
%filename 'figures/training_structure.eps';file-properties "XNPEU";}}}%
%BeginExpansion
%EndExpansion

\subsection{Channel estimation}

The matched filter in the Rake receiver in UWB systems is implemented using
analog circuits since it needs to operate at a high speed. The output of the
matched filter is sampled at symbol rate ($\mbox{1/T}_{\mbox{s}}=1/(p\Delta
))$. Therefore, during each symbol period, we can only observe $L$ outputs,
each from one of $L$ fingers. On the other hand, we need to estimate channel
coefficients every $\Delta$ seconds; thus we need to obtain $p$ uniform
samples during each symbol period.

In order to solve this seeming paradox, we use an approach that shows some
similarity to the \textquotedblleft swept time delay cross
correlator\textquotedblright\ channel sounder proposed in \cite{Cox_1972}. We
send the same training sequence (with guard interval) multiple times to obtain
denser sampling of the matched filter output. For a Rake receiver with $10$
fingers, $10$ samples with different timings can be obtained within one symbol
duration if the training sequence is sent once. Therefore, to get $32$ samples
per symbol duration, the training sequence needs to be repeated $4$ times (see
also Fig. \ref{training_structure}). Each training sequence consists of $511$
symbols, and $365$ ns guard interval to prevent interference caused by delay
spread of UWB channels between adjacent training sequences. Consequently, the
length of the whole training period for parameter estimation is $4(511\ast
5+365)=11600$ ns or $11.6$ $\mu$s. The detailed equations for the channel
estimates can be found in the Appendix.

Figs. \ref{channel_estimation_error1}, \ref{channel_estimation_error2} shows
the normalized MSE (NMSE) of our channel estimation, which is defined as%
\begin{equation}
NMSE=\frac{%
%TCIMACRO{\dsum \limits_{n}}%
%BeginExpansion
{\displaystyle\sum\limits_{n}}
%EndExpansion
|\widetilde{h}(n\Delta)-h(n\Delta)|^{2}}{%
%TCIMACRO{\dsum \limits_{n}}%
%BeginExpansion
{\displaystyle\sum\limits_{n}}
%EndExpansion
|h(n\Delta)|^{2}}\text{ \ .}%
\end{equation}
From Fig. \ref{channel_estimation_error1}, the channel estimation improves
with the signal-to-noise ratio when it is less than 25 dB. However, when it is
over 35 dB, there is an error floor. Fig. \ref{channel_estimation_error2}
shows the normalized MSE (NMSE) of the 10 largest channel taps, which is much
better than the NMSE of overall channel estimation.%
%TCIMACRO{\FRAME{ftbpFU}{15.0579cm}{5.0215cm}{0pt}{\Qcb{NMSE of overall channel
%estimation in IEEE 802.15.3a channel models (see Sec. VI).}}%
%{\Qlb{channel_estimation_error1}}{channel_estimation_error1.eps}%
%width 15.0579cm;  height 5.0215cm;  depth 0pt;  original-width 19.618cm;
%original-height 16.3231cm;  cropleft "0";  croptop "1";  cropright "2.5190";
%filename 'figures/channel_estimation_error1.eps';file-properties "XNPEU";}}}%
%BeginExpansion
\begin{figure}
[ptb]
\begin{center}
\includegraphics[
trim=0.000000cm 0.000000cm -29.799744cm 0.000000cm,
height=5.0215cm,
width=15.0579cm
]%
{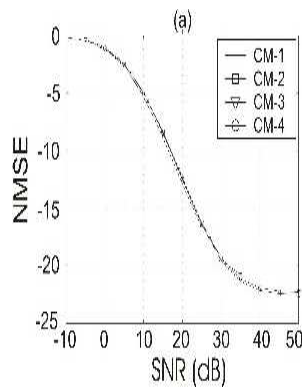}%
\caption{NMSE of overall channel estimation in IEEE 802.15.3a channel models
(see Sec. VI).}%
\label{channel_estimation_error1}%
\end{center}
\end{figure}
%EndExpansion%
%TCIMACRO{\FRAME{ftbpFU}{15.0579cm}{5.0215cm}{0pt}{\Qcb{NMSE of 10 largest
%channel taps in IEEE 802.15.3a channel models (see Sec. VI).}}%
%{\Qlb{channel_estimation_error2}}{channel_estimation_error2.eps}%
%width 15.0579cm;  height 5.0215cm;  depth 0pt;  original-width 19.372cm;
%original-height 16.2528cm;  cropleft "0";  croptop "1";  cropright "2.5402";
%filename 'figures/channel_estimation_error2.eps';file-properties "XNPEU";}}}%
%BeginExpansion
\begin{figure}
[ptbptb]
\begin{center}
\includegraphics[
trim=0.000000cm 0.000000cm -29.836754cm 0.000000cm,
height=5.0215cm,
width=15.0579cm
]%
{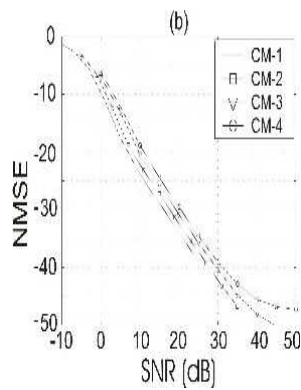}%
\caption{NMSE of 10 largest channel taps in IEEE 802.15.3a channel models (see
Sec. VI).}%
\label{channel_estimation_error2}%
\end{center}
\end{figure}
%EndExpansion

After having obtained the channel estimates, we determine the optimum Rake
combining weights by minimizing the mean-square error. The concatenation of
channel and Rake receiver constitutes a "composite" channel that is sampled
once per symbol. The equalizer is adapted such that it minimizes the
mean-square error of the equalizer output compared to a special training
sequence that is transmitted after the Rake weights have been adjusted.
Detailed equations about the weights for Rake and equalizer can be found in
the appendix.

\subsection{Synchronization}

Before any data demodulation can be done on the received UWB signal, the
template signal and the received signal must be time-aligned. The aim of
acquisition is to determine the relative delay of the received signal with
respect to the template signal. The conventional technique to achieve this is
the serial search algorithm. In this scheme, the received signal is correlated
with a template signal and the output is compared to a threshold. If the
output is lower than the threshold, the template signal is shifted by some
amount, which usually is comparable to the resolvable path interval and the
correlation with the received signal is obtained again. In this way, the
search continues until an output exceeds the threshold. If the output of the
correlation comes from a case where signal paths and the template signal are
aligned, it is called a signal cell output. Otherwise, it is called a
non-signal cell output. A false alarm occurs when a non-signal cell output
exceeds the threshold. In this case, time $t_{p}$ elapses until the search
recovers again. This time is called penalty time for false alarm.

However, in UWB systems, such a sequential search can be very time consuming,
as the number of cells is very large. This problem can be overcome by a new
algorithm that we call "sequential block search". The key idea here is to
divide the possible search space, which contains the cells, into several
blocks, where each of the blocks contains a number of signal cells. We then
first perform a quick test to check if the whole block contains a signal cell,
or not. Once we have identified the block that contains the signal, a more
detailed (sequential) search is performed in that block; for details, see
\cite{Gezici_et_al_2003}. Simulations show that acquisition can be achieved
(with 90\% probability) in less than $10\mu s$.\footnote{Note that the
treshold whether detection has taken place or not is a critical parameter of
the algorithm. A discussion of how to set this threshold can be found in
\cite{Gezici_et_al_2003}.} This can be shortened even further if the search
space is restricted, e.g., by exploiting knowledge from a beacon signal.

\section{Performance results}

In this section, we analyze the performance of our system in multipath and
interference. The performance of the system was simulated in \textquotedblleft
typical\textquotedblright\ UWB channels, which were developed within the IEEE
802.15.3a UWB standardization activities and are described in detail in
\cite{Molisch_et_al_2003_Mag}. We distinguish between four different types of
channels (called CM1, CM2, CM3, and CM4). CM1 describes line-of-sight (LOS)
scenarios with distances between TX and RX of less than $4$ m; CM2 and CM3
describe non-LOS scenarios at distances $0-4$, and $4-10$ m, respectively. CM4
is valid for heavy multipath environments. Note that in the following, we will
plot the performance in all the four different types of channels over a wide
range of distances. \footnote{We also evaluate the performance at distances
that the IEEE models were not originally intended for (e.g., CM1 was extracted
from measurements where the distance between TX and RX is less than 4m). We do
this as it gives insights into the relative importance of delay dispersion and
attenuation.}%

%TCIMACRO{\FRAME{ftbpFU}{8.0309cm}{5.0215cm}{0pt}{\Qcb{Probability of link
%success as function of distance for 110Mbit/s mode}}{\Qlb{link_success_100}%
%}{link_success_100.eps}{\special{ language "Scientific Word";
%type "GRAPHIC";  maintain-aspect-ratio TRUE;  display "USEDEF";
%valid_file "F";  width 8.0309cm;  height 5.0215cm;  depth 0pt;
%original-width 4.7496in;  original-height 3.2309in;  cropleft "0";
%croptop "1";  cropright "1.2622";  cropbottom "0";
%filename 'figures/link_success_100.eps';file-properties "XNPEU";}}}%
%BeginExpansion
\begin{figure}
[ptb]
\begin{center}
\includegraphics[
trim=0.000000in 0.000000in -1.245345in 0.000000in,
height=5.0215cm,
width=8.0309cm
]%
{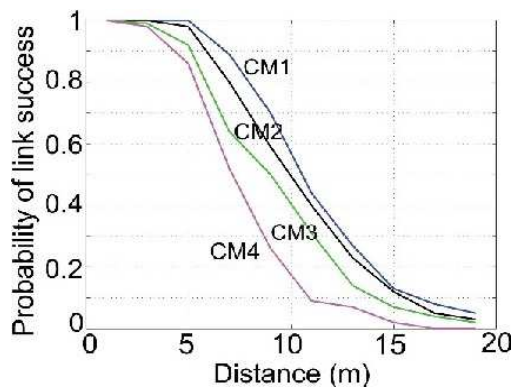}%
\caption{Probability of link success as function of distance for 110Mbit/s
mode}%
\label{link_success_100}%
\end{center}
\end{figure}
%EndExpansion

Fig. \ref{link_success_100} shows the probability for obtaining a successful
link. A \textquotedblleft successful\textquotedblright\ link means that
acquisition is obtained successfully, and the packet error probability (over
the ensemble of different channels) is less than 8{\%}. For CM1, the mean
coverage distance\footnote{The mean coverage distance is defined as the
distance where the packet error rate, averaged over all channel realizations,
is below the target rate.} is about $10$ m. The 10{\%} outage distance
(meaning that 8{\%} packet error rate or less is guaranteed in 90{\%} of all
channels) is $7$ m. For heavy multipath (CM4) these values decrease to $7$ and
$4$ m, respectively.%
%TCIMACRO{\FRAME{ftbpFU}{8.0309cm}{5.0215cm}{0pt}{\Qcb{Probability of link
%success as a function of distance for the 200Mbit/s mode.}}%
%{\Qlb{link_success_200}}{link_success_200.eps}%
%width 8.0309cm;  height 5.0215cm;  depth 0pt;  original-width 4.7496in;
%original-height 3.2309in;  cropleft "0";  croptop "1";  cropright "1.2700";
%filename 'figures/link_success_200.eps';file-properties "XNPEU";}}}%
%BeginExpansion
\begin{figure}
[ptb]
\begin{center}
\includegraphics[
trim=0.000000in 0.000000in -1.282392in 0.000000in,
height=5.0215cm,
width=8.0309cm
]%
{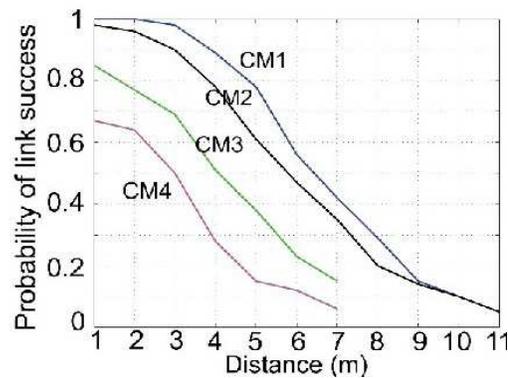}%
\caption{Probability of link success as a function of distance for the
200Mbit/s mode.}%
\label{link_success_200}%
\end{center}
\end{figure}
%EndExpansion

Fig. \ref{link_success_200} shows the analogous curves for data rate of $200$
Mbit/s. Due to the higher rate, the original data stream is converted
(demultiplexed) into two parallel data streams with $100$ Mbit/s each. The two
data streams are then transmitted simultaneously, using time hopping codes
that have the same hopping sequence, but are offset in delay by one chip. In
an AWGN channel, those codes would remain orthogonal, and the performance
should be worsened only by $3$ dB (since the E$_{b}$/N$_{0}$ is decreased).
However, in a multipath channel, the temporally offset codes lose their
orthogonality, which worsens the performance. One way to remedy this situation
is to use different (not just offset) hopping codes. However, this decreases
the number of possible simultaneous piconets. Another approach would be the
use of the scheme of \cite{Yang_and_Giannakis_2002}, which retains the
orthogonality of codes even in delay-dispersive channels.%

%TCIMACRO{\FRAME{ftbpFU}{8.0309cm}{5.0215cm}{0pt}{\Qcb{Packet error rate as a
%function of the distance of interfering piconet from the receiver (normalized
%to the distance between desired piconet transmitter to the receiver) in CM 1.
%}}{\Qlb{SOP_CM1}}{sop_cm1.eps}{\special{ language "Scientific Word";
%type "GRAPHIC";  maintain-aspect-ratio TRUE;  display "USEDEF";
%valid_file "F";  width 8.0309cm;  height 5.0215cm;  depth 0pt;
%original-width 4.7496in;  original-height 3.2309in;  cropleft "0";
%croptop "1";  cropright "1.2583";  cropbottom "0";
%filename 'figures/SOP_CM1.eps';file-properties "XNPEU";}}}%
%BeginExpansion
\begin{figure}
[ptb]
\begin{center}
\includegraphics[
trim=0.000000in 0.000000in -1.226821in 0.000000in,
height=5.0215cm,
width=8.0309cm
]%
{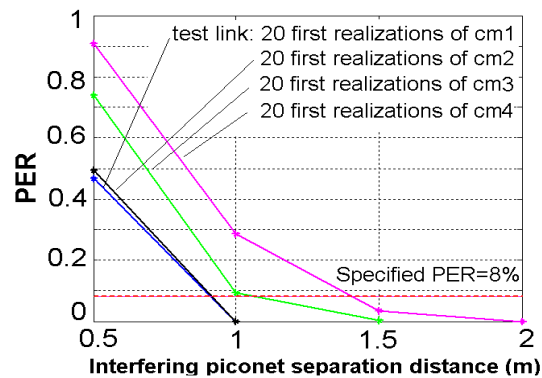}%
\caption{Packet error rate as a function of the distance of interfering
piconet from the receiver (normalized to the distance between desired piconet
transmitter to the receiver) in CM 1. }%
\label{SOP_CM1}%
\end{center}
\end{figure}
%EndExpansion%
%TCIMACRO{\FRAME{ftbpFU}{8.0309cm}{5.0215cm}{0pt}{\Qcb{Packet error rate as a
%function of the distance of interfering piconet in CM 3.}}{\Qlb{SOP_CM4}%
%}{sop_cm4.eps}{\special{ language "Scientific Word";  type "GRAPHIC";
%width 8.0309cm;  height 5.0215cm;  depth 0pt;  original-width 4.7496in;
%original-height 3.2309in;  cropleft "0";  croptop "1";  cropright "1.1313";
%cropbottom "0";  filename 'figures/SOP_CM4.eps';file-properties "XNPEU";}}}%
%BeginExpansion
\begin{figure}
[ptbptb]
\begin{center}
\includegraphics[
trim=0.000000in 0.000000in -0.623622in 0.000000in,
height=5.0215cm,
width=8.0309cm
]%
{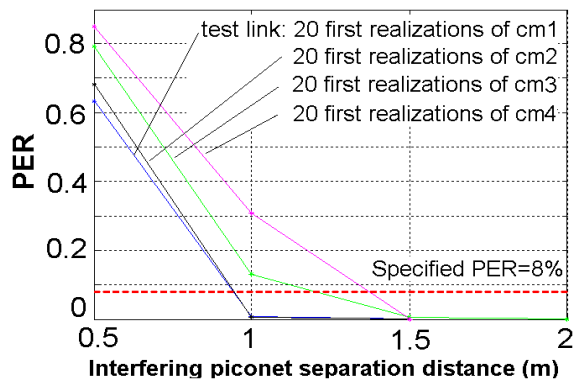}%
\caption{Packet error rate as a function of the distance of interfering
piconet in CM 3.}%
\label{SOP_CM4}%
\end{center}
\end{figure}
%EndExpansion

Figs. \ref{SOP_CM1} and \ref{SOP_CM4} show the performance when two users
(independent piconets) are operating simultaneously. The desired users are
located at half the distance that gives the 90{\%} outage probability (i.e.,
there is a $6$ dB margin\footnote{As the channel model prescribes the received
power to be proportional to $d^{-2}$, halving the distance means increasing
the power by $6$dB.} with respect to the single-user case); shadowing is not
considered in that graph. We find that an \textquotedblleft interfering
piconet\textquotedblright\ can be at a distance from the victim receiver of
about $1$ m (if the desired piconet is operating in CM1 or CM2), or $1.5$ m
(if the desired piconet is operating in CM3 or CM4). The performance does not
depend on which channel model is used for the interfering piconet.

Table 1 shows the coexistence of our system with other communications devices,
obeying various narrowband standards. In the column "desired", we list the
interference power that must not be exceeded according to the specifications
of the IEEE 802.15.3a technical requirements documentation (this power is
derived from receiver sensitivity specifications for various systems). In the
"achieved" column, we list the interference power (within the victim receiver
bandwidth) received from our UWB transmitter spaced at $1 $m distance from the
victim receiver. The column "FCC mask" gives the interference power created by
a UWB transmitter (at $1$m distance) that transmits at all frequencies with
the maximum power allowed by the FCC mask. We find that if the UWB transmitter
emits with the full power allowed by the FCC, it can significantly interfere
with other communications devices. A suppression of about 15dB is necessary to
allow coexistence within a 1m range. We achieve this suppression with the
spectral shaping as described in Sec. 3.3.

Finally, we also analyzed the resistance of the UWB system to interference
\textit{from} other communications devices. We found that again, a minimum
distance of 1m is sufficient to allow operation with less than $8{\%}$ PER.

\section{Summary and conclusions}

We have presented a UWB communications system based on time-hopping impulse
radio. This system uses only baseband components, while still being compatible
with FCC requirements, and providing a flexible shaping of the transmit
spectrum in order to accommodate future requirements by other spectrum
governing agencies, as well as not interfere with 802.11a wireless LANs and
other communications receivers in the microwave range. Our system can sustain
data rates of $110$ Mbit/s at $15$ m in AWGN channels, and $4-7$ m in
multipath channels. It is also resistant to interference from other UWB users,
as well as interference from wireless LANs, microwave ovens, and other interferers.

\begin{table}[ptbh]
\begin{center}%
\begin{tabular}
[c]{|p{53pt}|l|l|l|}\hline
\textit{System} & \textit{Desired} & \textit{Achieved} & \textit{FCC
Mask}\\\hline
$802.11a$ & \textit{-88dBm} & \textit{-90dBm} & \textit{-75dBm}\\\hline
$802.11b$ & \textit{-82dBm} & \textit{-85dBm} & \textit{-70dBm}\\\hline
$802.15.1$ & \textit{-76dBm} & \textit{-95dBm} & \textit{-80dBm}\\\hline
$802.15.3$ & \textit{-81dBm} & \textit{-85dBm} & \textit{-70dBm}\\\hline
$802.15.4$ & \textit{-91dBm} & \textit{-95dBm} & \textit{-80dBm}\\\hline
\end{tabular}
\label{tab1}
\end{center}
\end{table}

Table 1: Coexistence for other systems

\bibliographystyle{ieeetr}
\bibliography{art_ch1,uwb}

\section{Appendix A Parameter estimation}

To obtain uniform samples, the timing of the $l$-th finger corresponding to
the $m$-th training sequence is adjusted as follows:
\begin{equation}
t_{l,m}=4(l-1)\Delta+(m-1)\Delta,
\end{equation}
for $l=1,\cdots,10,$ and $m=1,\cdots,4.$

Let the training sequence be $b_{k}^{t}$'s for k=0, 1, {\ldots}, 510, where
superscript $^{t}$ denotes "training". Then the training signal can be
expressed as
\begin{equation}
s^{t}\left(  t\right)  =\sum\limits_{k=0}^{510}b_{k}^{t}w\left(
t-kT_{s}\right)  ,
\end{equation}

From Equation (\ref{eq10}), the $\Delta$-spaced output of the matched filter
is
\begin{equation}
y^{t}\left(  n\Delta\right)  =\sum\limits_{k=0}^{510}b_{k}^{t}\tilde{h}\left(
n\Delta-kp\Delta\right)  +\tilde{n}\left(  n\Delta\right)  .
\end{equation}
Consequently, the estimated channel taps can be expressed as%
\begin{equation}
\tilde{h}\left(  n\Delta\right)  =\frac{1}{511}\sum\limits_{k=0}^{510}%
b_{k}^{t}y^{t}\left(  n\Delta+kp\Delta\right)  .
\end{equation}

It can be shown that%
\begin{align}
\tilde{h}\left(  n\Delta\right)   & =h\left(  n\Delta\right)  +\frac{1}%
{511}\sum\limits_{m=0}^{510}\left(  \sum\limits_{k=m}^{510}b_{k}^{t}%
b_{k-m}^{t}\right)  h\left(  n\Delta+mp\Delta\right) \\
& +\frac{1}{511}\sum\limits_{m=-510}^{0}\left(  \sum\limits_{k=0}^{510-m}%
b_{k}^{t}b_{k-m}^{t}\right)  h\left(  n\Delta+mp\Delta\right)  +\frac{1}%
{511}\sum\limits_{k=0}^{510}b_{k}^{t}\tilde{n}\left(  n\Delta+kp\Delta\right)
\text{ \ .}\nonumber
\end{align}
The 2nd and 3rd terms in the above equation are the perturbations from other
taps due to imperfect orthogonality of the training sequence and the 4th term
presents the effect of channel noise.

To exploit the improved approach for UWB systems with co-channel interference,
interference power has to be estimated. Using the estimated channel and the
training sequence, the interference can be estimated by
\begin{equation}
i_{t}\left[  n\right]  =y_{t}\left(  n\Delta\right)  -\sum\limits_{k=0}%
^{510}b_{k}^{t}\tilde{h}\left(  n\Delta-kp\Delta\right)  ,
\end{equation}
and from it, interference-plus-noise power can be estimated by
\begin{equation}
P_{k}=\frac{1}{511}\sum\limits_{m=0}^{510}\left\vert i_{t}\left[  mp+k\right]
\right\vert ^{2},
\end{equation}
for $k=0,1,{\ldots},(p-1).$

Next, we determine the Rake weights. Let $n_{1},\cdots,n_{L}$ be the indices
of the $L$ largest taps. Then the weights for the MMSE Rake combiner and
optimum timing can be found by minimizing
\begin{equation}
MSE(\vec{\gamma},n_{0})=\frac{1}{511}\sum\limits_{n=0}^{510}\left\vert
z_{t}(n,n_{0})-b_{n}^{t}\right\vert ^{2}=\frac{1}{511}\sum\limits_{n=0}%
^{510}\left\vert \sum\limits_{l=1}^{10}\gamma_{l}y_{t}(pn+n_{l}+n_{0}%
)-b_{n}^{t}\right\vert ^{2}%
\end{equation}
Direct least-squares calculation yields that \cite{Haykin_1996}
\begin{equation}
\vec{\gamma}=\left(
{{\begin{array}{*{20}c} {\gamma _1 } \hfill \\ \vdots \hfill \\ {\gamma _{10} } \hfill \\ \end{array}}%
}\right)  =\left(  {\mathrm{\mathbf{Y}}_{t}\mathrm{\mathbf{Y}}_{t}^{H}%
}\right)  ^{-1}\left(  {\mathrm{\mathbf{Y}}_{t}\mathrm{\mathbf{b}}_{t}^{H}%
}\right)  ,
\end{equation}
where
\begin{equation}
\mathrm{\mathbf{Y}}_{t}=\left(
{{\begin{array}{*{20}c} {y_t \left[ {n_1 +n_o } \right]} \hfill & \hfill & \cdots \hfill & {y_t \left[ {510p+n_1 +n_o } \right]} \hfill \\ {y_t \left[ {n_2 +n_o } \right]} \hfill & \hfill & \cdots \hfill & {y_t \left[ {510p+n_2 +n_o } \right]} \hfill \\ \vdots \hfill & \hfill & \cdots \hfill & \vdots \hfill \\ {y_t \left[ {n_{10} +n_o } \right]} \hfill & \hfill & \cdots \hfill & {y_t \left[ {510p+n_{10} +n_o } \right]} \hfill \\ \end{array}}%
}\right)  ,
\end{equation}
and
\begin{equation}
\mathrm{\mathbf{b}}_{t}=\left(
{{\begin{array}{*{20}c} {b_0^t } \hfill & {b_1^t } \hfill & \cdots \hfill & {b_{510}^t } \hfill \\ \end{array}}%
}\right)  .
\end{equation}

From the estimated weights for the Rake receiver, its output can be calculated
by
\begin{equation}
z_{t}\left[  n,n_{o}\right]  =\sum\limits_{l=1}^{10}\gamma_{l}y_{t}\left[
pn+n_{l}+n_{o}\right]  .
\end{equation}
The equalizer coefficients can be estimated by minimizing
\begin{equation}
\frac{1}{511}\sum\limits_{n=0}^{511}\left\vert \sum\limits_{k=-L}^{L}%
c_{k}z_{t}[n-k,n_{o}]-b_{k}^{t}\right\vert ^{2}.
\end{equation}
Consequently \cite{Haykin_1996},
\begin{equation}
\left(
{{\begin{array}{*{20}c} {c_{-2} } \hfill \\ \vdots \hfill \\ {c_2 } \hfill \\ \end{array}}%
}\right)  =\left(  {\frac{1}{511}\sum\limits_{k=0}^{510}{\mathrm{\mathbf{z}%
}_{k}^{t}\mathrm{\mathbf{z}}_{k}^{t}{}^{T}}}\right)  ^{-1}\left(  {\frac
{1}{511}\sum\limits_{k=0}^{510}{\mathrm{\mathbf{z}}_{k}^{t}b_{k}^{t}}}\right)
,
\end{equation}
where
\begin{equation}
\mathrm{\mathbf{z}}_{k}^{t}=\left(
{{\begin{array}{*{20}c} {z_t \left[ {k+2,n_o } \right]} \hfill \\ \vdots \hfill \\ {z_t \left[ {k-2,n_o } \right]} \hfill \\ \end{array}}%
}\right)  .\label{eq12}%
\end{equation}

\end{document}